\documentclass[aps,preprint, groupaddress , amsmath, amssymb, showpacs, showkeys]{revtex4}
\usepackage{graphicx}
\usepackage{dcolumn}
\usepackage{bm}
\usepackage{upgreek}
\usepackage{times}
\usepackage{mathrsfs}

\begin{document}

\title{Single-carrier impact ionization favored by a limited band dispersion}
\author{A.~Darbandi}
\author{O.~Rubel}
\email[]{rubelo@tbh.net}
\affiliation{Thunder Bay Regional Research Institute, 290 Munro St, Thunder Bay, Ontario, Canada}
\affiliation{Lakehead University, 955 Oliver Road, Thunder Bay, Ontario, Canada}



\date{\today}

\begin{abstract}
A critical requirement for high gain and low noise avalanche photodiodes is the single-carrier avalanche multiplication. We propose that the single-carrier avalanche multiplication can be achieved in materials with a limited width of the conduction or valence band resulting in a restriction of kinetic energy for one of the charge carriers. This feature is not common to the majority of technologically relevant semiconductors, but it is observed in chalcogenides, such as Selenium and compound I$_2$-II-IV-VI$_4$ alloys.
\end{abstract}


\maketitle

%
%
\section{Introduction}\label{Sec:Introduction}

Impact ionization in semiconductors and subsequent avalanche multiplication of charge carriers is widely exploited in avalanche photodiodes (APDs), which have a wide range of applications in high-sensitivity photoreceptors \cite{Campbell_JLT_25_2007,David_IJSTiQE_14_2008,Lecomte_ITNS_43_1996}. APDs generally operate in a linear mode at a bias voltage lower than the breakdown voltage, where the output current is linearly proportional to the incident photon flux \cite{Yoshizawa2009}. The requirements to semiconducting materials for linear APDs include high gain and low excess noise. In order to fulfill both requirements, semiconductors that feature a single-carrier multiplication are needed. Unfortunately, the disparity between impact ionization coefficients for electrons and holes (so-called $K$-factor) is far from its ideal value for the majority of technologically-relevant semiconductors, such as Si \cite{Kasap_chapter2_Springer_2006}, GaAs (Ref.~\onlinecite{Adach1994}, p.~320), InP \cite{1}. In the following, we show that the single-carrier avalanche multiplication can be achieved as a result of a limited dispersion for one of the carriers and/or fundamental restrictions, such as energy and momentum conservation. This opens an avenue for the design of materials with the electronic structure that favors single-carrier avalanche multiplication.

\begin{figure}
	\includegraphics[width=0.4\columnwidth]{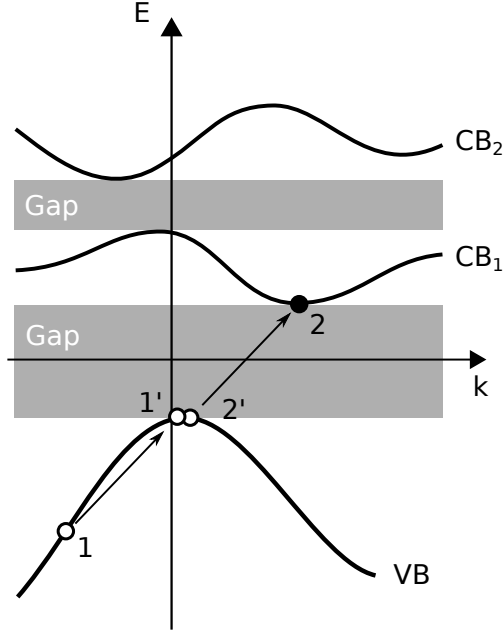}\\
	\caption{Hypothetical band structure of a material with frustrated avalanche multiplication of electrons. Successful ionization event is caused by the primary hole 1 and results in creation of the electron-hole pair $2-2'$. The distinctive feature of this band structure is the presence of an energy gap between conduction bands CB$_1$ and CB$_2$. The kinetic energy gained by the electron is then fundamentally limited by the width of CB$_1$ band.}\label{Fig:fig1}
\end{figure}

The hypothetical electronic band structure of a material with high $K$-factor is shown on Fig.~\ref{Fig:fig1}. Its characteristic feature is a limited width of the conduction band, which prevents free electrons from gaining the excess energy enough to initiate the impact ionization. In contrast to free electrons, the valence band dispersion permits the impact ionization initiated by holes. We anticipate that the material with such an electronic structure will exhibit high gain and full suppression of the counterpart carrier multiplication.

In order to illustrate feasibility of the proposed concept, we present a detailed analysis of the electronic structure of trigonal Selenium (t-Se) in the framework of a density functional theory (DFT). It is shown that the presence of an energy gap in the conduction band limits the electron participation in the impact ionization, while the extended dispersion of lone-pair states in the valence band favors the hole impact ionization.  Among compound semiconductors, group I$_2$-II-IV-VI$_4$ chalcogenides exhibit a unique electronic structure that fosters single-carrier multiplication.

%
%
\section{Method}\label{Sec:Method}

A necessary condition for the impact ionization to occur is that all initiating and resultant particles should satisfy the energy and momentum conservation requirements
	\begin{eqnarray}
		E(\mathbf{k}_{1},n_{1}) + E(\mathbf{k}_{2},n_{2}) &=& E(\mathbf{k}_{1'},n_{1'}) + E(\mathbf{k}_{2'},n_{2'})~,\label{Eq:Energy-conserv}\\
		\mathbf{k}_{1} + \mathbf{k}_{2} &=& \mathbf{k}_{1'} + \mathbf{k}_{2'}~,\label{Eq:Momentum-conserv}
	\end{eqnarray}
where $\mathbf{k}_i$ is the electron wave vector, $n_i$ is the band index and $E(\mathbf{k}_i,n_i)$ is the corresponding energy eigenvalue. The indices 1,~2 and $1'$,~$2'$ represent the initiating and resultant carriers, respectively. The ionization energy corresponds to the excess energy of the primary carrier~1.

Further analysis of the ionization threshold for electrons and holes in \mbox{t-Se} requires detailed knowledge of the band structure, which can be obtained self-consistently using DFT and an equilibrium atomic structure. The structure of t-Se consists of parallel helical chains arranged with a hexagonal symmetry, which can be characterized by two lattice constants and the radius of the chain \cite{2}. The full structural optimization was performed using planewave method implemented in the \texttt{ABINIT} package \cite{3,4}, generalized density approximations \cite{5} and Troullier-Martins pseudopotential \cite{6,7}. Convergence tests were performed with respect to the $k$-mesh density and the plane wave cutoff energy $E_{cut}$. The convergence was reached at $E_{cut}=25$~Ha and $4\times4\times4$ Monkhorst-Pack $k$-point mesh \cite{8}. The deviation between theoretical and experimental structural parameters did not exceed 4\%~\cite{9}.

\begin{figure*}
	\includegraphics[width=\textwidth]{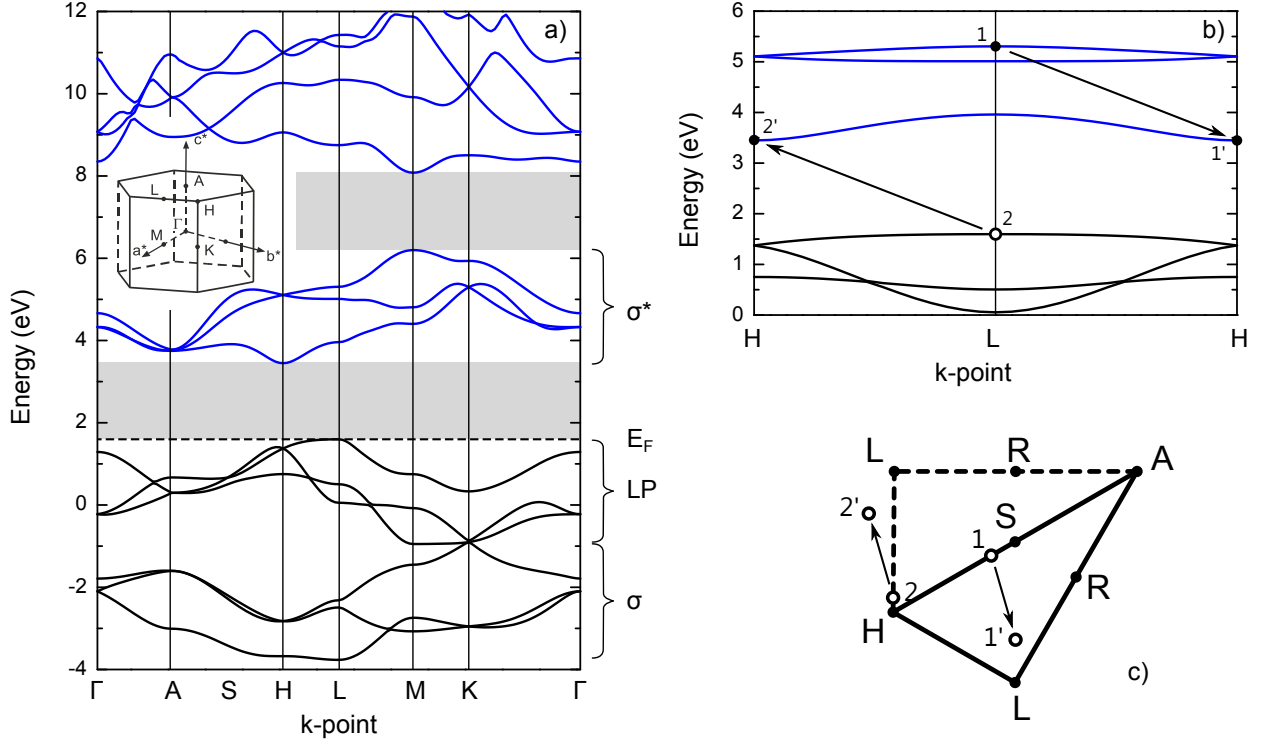}\\
	\caption{Band structure of t-Se (a) along high-symmetry points in the Brillouin zone, which is shown in the inset. Bonding $\sigma$, antibonding $\sigma^*$ and lone-pair (LP) states are indicated. The shaded areas corresponds to the optical gap and the gap between two conduction bands (from bottom to top, respectively). Panel (b) shows the band structure along $H-L-H$ segment with four particles involved into the electron ionization event near to the threshold. The top view of the Brillouin zone with k-points involved into the hole ionization event near to the threshold is illustrated on panel (c).}\label{Fig:fig2}
\end{figure*}

The calculated band structure of \mbox{t-Se} is presented in Fig.~\ref{Fig:fig2}. Trigonal Selenium is an indirect semiconductor with the lowest energy transition between $L$-point in the top of valence band and $H$-point in the bottom of conduction band. The DFT energy gap is about 1~eV, which is significantly underestimated with respect to the experimental value of 1.85~eV \cite{10,11,12}. This inconsistency is attributed to a well-known shortcoming of explicit density-dependent functionals, which tend to underestimate the energy gap \cite{13}. In Fig.~\ref{Fig:fig2} and the following analysis, the so-called ``scissor operator" (energy offset) was applied in order to match the theoretical energy gap with its experimental value.

The necessary precursor for impact ionization is a low threshold energy of a primary charge carrier. In the search for the ionization threshold, we sample the entire Brillouin zone using 6,400 k-points ($20\times20\times16$ mesh). Next we analyze all possible ionization events by generating combinations of four k-points that obey Eq.~(\ref{Eq:Momentum-conserv}). For each set of k-points and the combination of four bands ($n_{1}$, $n_{2}$, $n_{1'}$, $n_{2'}$) involved in the impact ionization event, the energy conservation criterion Eq.~(\ref{Eq:Energy-conserv}) is evaluated using a Gaussian approximation for the delta function with the smearing of 50~meV. This approach provides the resolution of approximately 20~meV for the ionization energy, which is sufficient for the purpose of our discussion. The threshold energy is identified as the lowest possible ionization energy that satisfies the conservation rules.

%
%
\section{Results and discussion}

The analysis of the electronic structure of t-Se yields the ionization threshold of 1.85~eV for electrons as primary charge carriers. The charge carriers with the lowest ionization energy are located along $H-L-H$ segment of the Brillouin zone (see Fig.~\ref{Fig:fig2}b). The resultant carriers $1'$, $2$ and $2'$ occupy states in the valence band maximum (L-point) and the conduction band minimum (H-point), which ensures the lowest possible ionization energy $E_\text{th,e}= E_\text{g}$, providing a sharp contrast with the parabolic band approximation \cite{14}, which predicts $E_\text{th}=1.5E_g$.

Results of our calculations suggest that the ionization energy for primary holes amounts to 1.95~eV ($E_\text{th,h}=1.05 E_\text{g}$), which is slightly greater than that for electrons. The interpretation of this result requires further analysis of the valence band dispersion. For the energy threshold to be equal to the band gap, the resultant carriers $1'$, $2$ and $2'$ should not have any excess energy, i.e., occupy the valence band maximum and the conduction band minimum ($L$ and $H$ points, respectively). In this case, the momentum conservation dictates that the wave vector of the primary carrier $\mathbf{k}_{1}$ should be either at $H$ or $S$ point (see Fig.~\ref{Fig:fig2}c). Apparently, the excess energy of holes in $H$ and $S$ points of the lone pair band does not exceed the band gap (see Fig.~\ref{Fig:fig2}a), which precludes their participation in the impact ionization. Therefore, the wave vectors of particles participating in the impact ionization of holes near to the threshold deviate from these high symmetry points as shown in Fig.~\ref{Fig:fig2}c, resulting in the value of ionization energy slightly higher than $E_\text{g}$, but still lower than $1.5E_\text{g}$.

The low ratio $E_\text{th}/E_g\approx1$ is a consequence of an indirect band structure inherent to t-Se, which is a positive factor for the avalanche gain. However, the insufficient disparity between the values of $E_\text{th,e}$ and $E_\text{th,h}$ can affect the feasibility of operation in the single-carrier multiplication mode. It has been indeed observed experimentally \cite{15} that the avalanche multiplication in polycrystalline t-Se is dominated by holes with the ratio $\beta/\alpha\gg10$. This result implies that factors other than the ionization threshold are responsible for the nearly single-carrier multiplication in t-Se.

The impact ionization rate for a carrier with the energy equal to or above the ionization threshold is proportional to the value of a corresponding Auger matrix element and an effective density of states. In the following, we focus our discussion on the energy dependence of the effective density of states (DOS), assuming that the matrix element does not vary significantly near to the threshold energy \cite{16}. The effective DOS for four-particles inverse Auger process represents the relative number of possible ionization events as a function of excess energy of the primary carrier.

\begin{figure}
	\includegraphics[width=0.6\columnwidth]{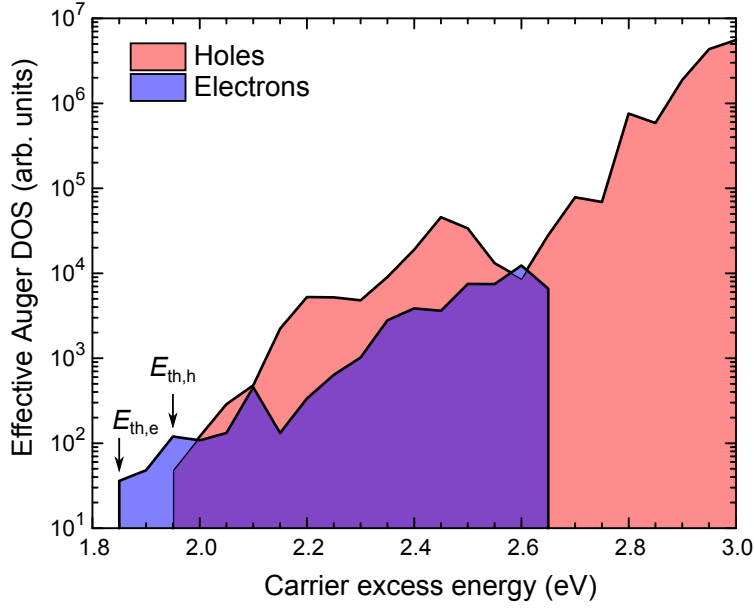}\\
	\caption{Effective density of states for impact ionization as a function of the excess energy of primary electrons and holes.}\label{Fig:fig3}
\end{figure}

The relative number of possible ionization events that satisfy the conservation rules is plotted on Fig.~\ref{Fig:fig3} as a function of the excess energy of the primary carrier. Apparently, the effective DOS for primary holes increases faster than that for electrons as the excess energy exceeds the threshold. Furthermore, the DOS for primary electrons exists only in the narrow range of excess energies $1.85-2.65$~eV, which is related to a limited width of $\sigma^*$ antibonding states in t-Se. The effective DOS for holes shows no discontinuity at higher excess energies due to the absence of the energy gap between lone-pair and bonding $\sigma$ states as can be seen at Fig.~\ref{Fig:fig2}a.

The results of our calculations suggest that the suppression of electrons' avalanche multiplication in t-Se can be caused by peculiarities of its electronic structure. In conjunction with the week electron-phonon coupling inherent to holes in t-Se \cite{17,9}, these properties favor the single-carrier impact ionization.

Amorphous Selenium also reveals a high disparity between impact ionization coefficients for electrons and holes. The $K$-factor for amorphous selenium ranges between 10 and 100 depending on the electric field strength \cite{18}. However, the field corresponding to the onset of avalanche multiplication in a-Se is about 70~V/$\mu$m \cite{19}, which is significantly higher than that in the polycrystalline form ($15\ldots20$~V/$\mu$m). This difference is attributed to intense elastic scattering due to disorder inherent to the amorphous structure \cite{20,21}.

Another prominent class of materials that demonstrated a well-separated narrow conduction band similar to Fig.~\ref{Fig:fig1} are quaternary chalcogenide I$_2$-II-IV-VI$_4$ semiconductors \cite{22,23,24}. Specific examples include Cu$_2$ZnGeS$_4$ and Cu$_2$ZnSnSe$_4$ compounds, which are currently studied as an active material for thin-film solar cells \cite{25}. In these structures, the lowest conduction band has the width of $1-2$~eV, and it is separated by the energy gap of approximately the same magnitude from the rest of the conduction band \cite{22,23,24}. The topmost valence band, on the other hand, is much wider than the optical energy gap. Bearing in mind that the optical energy gap in these materials is of the order of 1.5~eV, this material class show a great promise for single-carrier (holes) avalanche multiplication, because the electron multiplication should be fully suppressed.

\section{Conclusions}

The single-carrier multiplication is a requirement for the development of linear avalanche photodiodes with high gain and low noise. We show that the single-carrier multiplication regime can be achieved by confinement of the kinetic energy for the counterpart charge carrier. This hypothesis is demonstrated through a detailed analysis of the electronic structure of trigonal Selenium (t-Se) in the framework of a density functional theory. The ionization threshold energies for electrons and holes in t-Se were computed based on its electronic structure. Our results suggest that the excess energy of the primary carrier required to generate the secondary electron-hole pair in t-Se is approximately equal to the energy gap ($E_\text{th,e/h}\approx E_\text{g}$) that is significantly different from the result predicted by the parabolic band approximation ($E_\text{th,e/h}=1.5 E_\text{g}$). The effective density of states for the impact ionization with holes as a primary carrier exceeds by orders of magnitude the corresponding value for electrons. This result is due to the presence of an energy gap in the conduction band that limits electrons from acquiring high kinetic energy, which is not the case for holes. The latter favors the high disparity between impact ionization coefficient of holes and electrons, which is consistent with the experimental observation of dominant holes avalanche multiplication in poly-crystalline Selenium. Similar peculiarities of the electronic band structure are found in compound chalcogenide I$_2$-II-IV-VI$_4$ semiconductors that makes them a promising candidate for the active material in high-sensitivity photoreceptors.

\begin{acknowledgments}
Financial support of Natural Sciences and Engineering Research Council of Canada under a Discovery Grants Program ``Microscopic theory of high-Þeld transport in disordered semiconductors", Ontario Ministry of Research and Innovation through a Research Excellence Program ``Ontario network for advanced medical imaging detectors" and Thunder Bay Community Economic and Development Commission is highly acknowledged.
\end{acknowledgments}

%
%

\end{document}